# Supersymmetry and PT-Symmetric Spectral Bifurcation

Kumar Abhinav† and Prasanta K. Panigrahi*,

IISER-Kolkata, Mohanpur, Nadia-741252, West Bengal

[ To be published in "Photonics and Quantum Structures": Proceedings of national Seminar on Photonics and Quantum Structures-2009]

**Abstract:**

Dynamical systems exhibiting both PT and Supersymmetry are analyzed in a general scenario. It is found that, in an appropriate parameter domain, the ground state may or may not respect PT-symmetry. Interestingly, in the domain where PT-symmetry is not respected, two superpotentials give rise to one potential; whereas when the ground state respects PT, this correspondence is unique. In both scenarios, supersymmetry and shape-invariance are intact, through which one can obtain eigenfunctions and eigenstates exactly. Our procedure enables one to generate a host of complex potentials which are not PT-symmetric, and can be exactly solved.

## Introduction

Complex potentials symmetric under simultaneous Parity (P) and Time-reversal (T) transformations are known to yield real energy eigenvalues in suitable parameter domain, sharing common eigenfunctions with the PT operator [1-3]. Beyond this range, these PT-

Electronic addresses: †kumarabhinav@iiserkol.ac.in, *pprasanta@iiserkol.ac.in

symmetric potentials correspond to complex-conjugate spectra with eigenfunctions connected through PT operation. This ***spontaneous breaking*** of PT-symmetry is expected as PT is a non-linear operator. Various models have been studied, both numerically and analytically, to illustrate the above structure [4, 5]. When these systems are analyzed under Supersymmetric quantum mechanics (SUSY-QM) [6], in the real domain, they are found to be isospectral to a real potential [7]. We here put forward an approach to arrive at both real and complex-conjugate domains starting from a class of superpotentials, with complex parameters. For the real case, it is possible to have a unique superpotential. For the complex-conjugate domain we have multiple superpotentials, each representing a ***separate*** section of the complete Hilbert space, but possessing the same potential. Moreover, these two classes of superpotentials are related through simple parametric variation, showing a bifurcation in the complex energy plane.

In the following, we will discuss SUSY-QM briefly and then directly proceed to discuss a particular example of PT-symmetric potential [8]. Analysis of other complex potentials which are not PT-symmetric will be discussed, followed by some recent experimental findings.

## 1. SUSY-QM: A Brief Introduction

Following the factorization approach to analytically solvable Hamiltonians [9], in SUSY-QM [6, 10, 11, 12], the Hamiltonian can be written as $H_-(x) = A^\dagger A$, where $A^\dagger = -\partial/\partial x + W(x)$ and $A = \partial/\partial x + W(x)$; with the ***superpotential*** $W(x)$ being defined as: $V_\pm(x) = W^2(x) \pm \partial W/\partial x$. Here $V_-$ is the potential corresponding to $H_-(x)$ and $V_+$ corresponds to another Hamiltonian $H_+(x) = AA^\dagger$, which is the ***superpartner*** of $H_-(x)$. The eigenstates of $H_\pm(x)$ have one-to-one correspondence:

$$\psi_n^+(x) = [E_{n+1}^-]^{-1/2} A \psi_{n+1}^-(x),$$

and,

$$\psi_{n+1}^-(x) = [E_n^+]^{-1/2} A^\dagger \psi_n^+(x); \tag{1.1}$$

except for the ground state of $H_-(x)$, which is defined as $A\psi_0(x) = 0$, and is expressible as

$$\psi_0(x) = e^{-\int^x W(x')dx'}. \tag{1.2}$$

Also, from (1.1), the energies are related as $E_n^+ = E_{n+1}^-$, whence the potentials $V_\pm(x)$ are called ***isospectral***. Further, Gendenshtein [13] showed that, if two isospectral potentials are related as $V_+(x; a_0) = V_-(x; a_1) + R(a_1)$, where $a_0$ is a parameter in $V_\pm$, $a_1 = f(a_0)$ and $R(a_1)$ is independent of $x$; then the potentials are called ***shape-invariant***. From such potentials, one can construct a hierarchy of Hamiltonians:

$$H^s = -\partial^2/\partial x^2 + V_-(x; a_s) + \sum_{k=1}^{s} R(a_k) = -\partial^2/\partial x^2 + V_+(x; a_{s-1}) + \sum_{k=1}^{s-1} R(a_k),$$

with ground state energy $E_0^s = \sum_{k=1}^{s} R(a_k)$. On identifying, $H^1 = H_+$ and $H^0 = H_-$, the energy of the *n*th level of $H_-$ is found to be $E_0^n = \sum_{k=1}^{n} R(a_k)$. Further, the excited states of the original Hamiltonian can be found from the ground state of $H^n$ as:

$$\psi_n^-(x; a_s) \propto A^\dagger(x; a_0) A^\dagger(x; a_1) \ldots\ldots\ldots\ldots\ldots\ldots\ldots\ldots A^\dagger(x; a_{n-1}) \psi_0^-(x; a_n). \tag{1.3}$$

Therefore if the potential is shape-invariant, the eigenvalue problem can be solved completely by this algebraic method and the spectrum with corresponding eigenfunctions can be obtained. The application of the SUSY-QM approach to PT-symmetric pseudo-Hermitian Hamiltonians has been carried out in detail [14, 15], including the construction of the appropriate norm. Recently [16], the procedure to construct the norm for general pseudo-Hermitian Hamiltonians has been shown.

## 2. Construction of Complex PT-symmetric Potentials

Non-Hermitian, complex PT-symmetric potentials are known to have real eigenvalues over a definite range of parameters appearing in them, beyond which they show ***complex-conjugate*** (CC) spectra, with corresponding wavefunctions being PT-related. If such potentials are shape-invariant then the eigenvalue problem can be solved through SUSY-QM.

We start with general superpotentials with complex parameters, leading to both real and CC spectra for different range of the parameter values. In the real domain, for a given potential, the superpotential is ***unique***, for a given parameter range. For the CC domain, two ***different*** superpotentials yield the ***same*** potential, for a different range of parameterization. Thus, we arrive at the SUSY condition for ***phase-transition*** of the spectrum from real to CC values owing to spontaneous PT-symmetry breaking.

For example, we will consider generalizations of PT-symmetric shape-invariant complex potentials, $V(x;\alpha) = -V_1 sech^2(\alpha x) - iV_2\,\text{sech}(\alpha x)\tanh(\alpha x)$, which was analytically solved by Z. Ahmed [8]. Here $V_{1,2}$ and $\alpha$ are real constant parameters. We propose a superpotential,

$$W_{PT}^{\pm} = (A \pm iC^{PT})\tanh(\alpha x) + (\pm C^{PT} + iB)\,\text{sech}(\alpha x), \tag{2.1}$$

with $A, B, \alpha$ are real parameters. The corresponding potentials are,

$$V_{-}^{\pm}(x) = -[(A \pm iC^{PT})(A \pm iC^{PT} + \alpha) - (\pm C^{PT} + iB)^2]sech^2(\alpha x)$$
$$-i(\pm C^{PT} - B)[2(A \pm iC^{PT}) + \alpha]\,\text{sech}(\alpha x)\tanh(\alpha x). \tag{2.2}$$

This potential is not PT-symmetric, for which the co-efficient of the even-P term has to be real and that of the odd-P part has to be imaginary. Invoking this condition, we arrive at the ***bifurcation condition***,

$$C^{PT}[2(A - B) + \alpha] = 0. \tag{2.3}$$

From (2.3), there can be two possibilities. For $C^{PT} = 0$, we have from (2.1),

$$W_{PT}(x) \equiv W_{real}(x) = A\tanh(\alpha x) + iB\,\mathrm{sech}(\alpha x), \tag{2.4}$$

giving, from (2.2),

$$V_-^{\pm}(x) \equiv V_-(x) = -[A(A+\alpha) + B^2]sech^2(\alpha x) + iB(2A+\alpha)\mathrm{sech}\,(\alpha x)\tanh\,(\alpha x). \tag{2.5}$$

Then, from (2.4), we get the ***real*** spectrum of the above potential through shape-invariance as,

$$E = -(n\alpha - A)^2, \tag{2.6}$$

modulo a constant term, and from (1.3), the eigenfunctions as [9],

$$\psi_n(x) \propto [\sec\,(\alpha x)]^{\frac{A}{\alpha}} \exp\left[-i\frac{B}{\alpha}tan^{-1}\{sinh(\alpha x)\}\right] P_n^{-\frac{A}{\alpha}-\frac{B}{\alpha}-\frac{1}{2},-\frac{A}{\alpha}+\frac{B}{\alpha}-\frac{1}{2}}[i\sinh\,(\alpha x)], \tag{2.7}$$

Again, if $C^{PT} \neq 0$, from (2.3), one gets $A = B - \frac{\alpha}{2}$, which when substituted in (2.1) and (2.2), yields,

$$W_{PT}^{\pm}(x) \equiv W_c^{\pm}(x) = (A \pm iC^{PT})\tanh(\alpha x) + \left[\pm C^{PT} + i\left(A + \frac{\alpha}{2}\right)\right]\mathrm{sech}\,(\alpha x), \tag{2.8}$$

and, $$V_-^{\pm}(x) \equiv V_-^c(x) = -\left[2A(A+\alpha) - 2(C^{PT})^2 + \frac{\alpha^2}{4}\right]sech^2(\alpha x) + i\left[2A(A+\alpha) + 2(C^{PT})^2 + \frac{\alpha^2}{2}\right]\mathrm{sech}(\alpha x)\tanh(\alpha x). \tag{2.9}$$

Corresponding to two different superpotentials, we arrive at the CC spectrum,

$$E_n^{\pm} = 2n(A \pm iC^{PT})\alpha + (n\alpha)^2, \tag{2.10}$$

modulo a constant term again, and the eigenfunctions as,

$$\psi_n^{\pm}(x) \propto [\mathrm{sech}\,(\alpha x)]^{\frac{1}{\alpha}(A \pm iC^{PT})} \exp\left[\left\{-\frac{1}{\alpha}\left(A + \frac{\alpha}{2}\right) \mp \frac{C^{PT}}{\alpha}\right\} tan^{-1}\{sinh(\alpha x)\}\right] \times P_n^{\pm i2\frac{C^{PT}}{\alpha},2\frac{A}{\alpha}+\frac{1}{2}}[i\sinh\,(\alpha x)]. \tag{2.11}$$

The specific parameterization condition $C^{PT} \neq 0$ is the SUSY criterion for spontaneously broken PT, which is different from the analytic parameter criterion with PT being unbroken over a ***range*** of parameters. Further, for broken PT, two superpotentials corresponding to the ***bifurcation*** in the Hilbert space results into the same complex PT-symmetric potential. This non-uniqueness is different from isospectral deformation as it arises ***parametrically***. In the complex case shape-invariance leads to complex energy shifts, while in the real case this shift is real, which was observed earlier [17]. The key structure of such potentials is that the odd Parity part has purely imaginary co-efficient and the even part has purely real co-efficient. More examples are listed in the source paper.

## Conclusion

One can construct non-PT-symmetric shape-invariant complex potential through ***minimal complexification*** of the superpotential corresponding to the real counterparts []. For example, the Pöschl-Teller potential [18] $U(x) = U_a sech^2(\alpha x) + U_b csch^2(\alpha x)$, where $U_{a,b}$ and α are constant parameters, can be complexified by considering the superpotentials $W_1 = Atanh(\alpha x) + iBcoth(\alpha x)$ or $W_2 = iAtanh(\alpha x) + Bcoth(\alpha x)$, $A, B$ and α being constant parameters. In both cases, the spectra are complex. The non-uniqueness of the real spectra for the real potential is lifted by the complexification, with the imaginary part being equispaced. Further, the wave-functions become normalizable over a greater parameter range. A complex radial Coulomb potential can be constructed from the superpotential, $W(r) = \frac{i\alpha}{r} + \beta$, α and β being independent of *r*, which results into complex principal quantum numbers, with the wave-function being normalizable over a greater parameter range than those corresponding to the real counterpart [19]. Recently, spontaneous PT breaking has been observed experimentally in optical fibers, where the gain-loss profile showed bifurcation in the complex plane beyond a critical value of the ***optical loss co-efficient*** [20]. We propose similar observation in terms of our parameters, if they are identified properly for an optical system.

\# Recently it has been shown for unbroken PT-symmetry, the potential in Eq.(2.5) corresponds to not one, but two superpotentials, under the additional sl(2) symmetry of the system [21,22]. But each of them is shown to be independently mapped to the same pair of superpotentials in the broken PT sector under the SUSY parameterization [23].

**Acknowledgement**: We would like to acknowledge Prof. R. Dutt for his invaluable suggestions, and Profs. A. Mostafazadeh and M. Plyushchay for bringing a number of related references to our notice.